# UBV CCD standards near 5C1 radio sources at intermediate galactic latitude - I

V.Mohan[1], D.K.Ojha[2,3], O.Bienaymé[2], D.C.Paliwal[1], A.K.Pandey[1], A.C.Robin[2,3], M.Chareton[2], B.C.Bhatt[1], B.B.Sanwal[1], E.Oblak[2], M. Haywood[2] and M.Crézé[3]

[1] *Uttar Pradesh State Observatory, Manora Peak, Nainital, 263129, India*
[2] *Observatoire de Besançon, 41 bis, Av de l'Observatoire, BP1615, F-25010 Besançon Cedex, France*
[3] *Observatoire de Strasbourg, CNRS URA 1280, 11 rue de l'Université, F-67000 Strasbourg, France*

## Abstract

Using CCD observations photometric magnitudes have been obtained in 18 subfields in a direction at intermediate galactic latitude ($l = 170°$, $b = 45°$) in a 14 deg$^2$ area. B and V magnitudes have been obtained for a total of 214 stars out of which for 73 stars U magnitudes also have been obtained. The magnitude range covered is $11 < V < 21$ and $0.18 < B-V < 1.85$.

**keywords :**  CCD photometry – calibration – stellar populations

## 1. Introduction

We are having an on going programme of star counts on Schmidt plates in a few selected directions in the Galaxy with an aim to study galactic stellar populations. In order to calibrate Schmidt plates photometrically, it is necessary to have a number of photometric standards which cover the entire range of magnitudes to be studied on the plates. As Schmidt plates cover a wide area of the sky, the standards should also be spread over the surface of the plate so as to minimise geometrical effects present on the plate. This paper is first in a series of papers aimed at to provide photometric standards in the selected galactic directions to be studied. In this paper we present photometric magnitudes obtained through CCD observation in the galactic direction centered at $\alpha_{2000} = 9^h 41^m 20^s$ and $\delta_{2000} = +49°54'20''$ and covering an area of 14deg$^2$. The observed stars lie in the magnitude range 11<V<21 and 0.18<B-V<1.85. Notni et al. (1979) have also observed a few photoelectric and photographic standards in the same 5C1 direction.

## 2. Observations

The observations were obtained, using CCD systems at the f/13 Cassegrain focus of the 104-cm telescope of the Uttar Pradesh State Observatory (UPSO) in India and at f/6 Newtonian focus of the 120-cm telescope of the Observatoire de Haute-Provence (OHP) in France during January 1991 to January 1993.

At UPSO the observations were obtained in UBV passbands using a Photometrics CCD system which has a coated Thomson chip having 23 micron square 384 × 576 pixels. The chip covers an area of 2' × 3' at the Cassegrain focus of the 104-cm telescope of the Observatory. The details of the UPSO CCD system and filters used have been described by Mohan et al. (1991). At OHP the observations have been obtained in B and V passbands using a CCD system consisting of a RCA back illuminated chip having 30 micron square 323 × 512 pixels. Each pixel corresponds to 0.83" at the Newtonian focus of the 1.2 metre telescope and the toal field obtained is 4.4' × 7'. The details of the CCD system and the filters used have been given in Chevalier and Ilovaisky (1991).

We have obtained CCD observations of 18 sub-fields in this region. The total exposure time varies from 20 minutes in V to 60 minutes in U filters. The log of observations is given in table 1 and the identification charts are given on figure 1.

At UPSO, we have sandwiched the exposures of each field in each filter with exposure of a comparison field (field 12) in the same filter. The comparison field was same for all the observed fields and lies in the region of observations. The observations were carried out within two hours of the meridian so that air mass change within a given sandwich was only marginal. The comparison field was standardised using Landolt's stars (1983) on two different nights.

At OHP, the observations were taken in B and V passbands for a few sub-fields without any comparison field being observed. A sub-portion of each field observed at OHP was re-observed at UPSO using the above mentioned procedure, so as to tie the OHP magnitudes to the standard system. A number of bias and twilight flat-field frames were also taken on several nights during the observing runs both at UPSO and OHP.

### 3. Data reduction

The data analysis was mainly carried out using the computing facilities of Observatoire de Besançon. The frames were cleaned employing the standard procedures using ESO MIDAS software running on VAX and DEC stations of the Observatory. Different clean frames of same field in same filter were co-added. Photometry on thus obtained frames was carried out using DAOPHOT photometric package by Stetson (1987). PSF was obtained for each frame and the PSF magnitudes were suitably tied to aperture photometry magnitudes. For comparison field only aperture photometry was obtained.

There were no saturated stars in the UPSO observations, however, a few stars were saturated in long exposure frames of the OHP observations. For such stars photometry has been carried out using unsaturated short exposure frames only.

For each frame observed at UPSO, differential magnitudes and colours were obtained using the observed magnitudes of the comparison field. These differential magnitudes were then standardised using the transformation equations obtained by Mohan et al. (1991) for the UPSO system :

$$\Delta(V - v) = -0.018\Delta(B - V) \quad (1)$$

$$\Delta(B - V) = 1.192\Delta(b - v) \quad (2)$$

$$\Delta(U - B) = 0.918\Delta(u - b) \quad (3)$$

For OHP observations , the sub-sets of stars standardised at UPSO were used in conjunction with the transformation equations obtained by Chevalier and Ilovaisky (1991) for the OHP CCD system :

$$(V - v) = 0.043(b - v) + \beta 1 \tag{4}$$

$$(B - V) = 1.104(b - v) + \beta 2 \tag{5}$$

where $\beta 1$ and $\beta 2$ were determined using the standards in the frame obtained at UPSO. The precision of photometric measures as determined by comparing different frames varies from 0.02 mag. for V = 12 mag. to 0.10 mag. for V = 20 mag.

In table 2 we have listed the identification numbers, co-ordinates (epoch 2000), V, (B-V) and (U-B) for each star observed by us. A total of 214 stars have been observed. While all the stars have V and (B-V) magnitudes, (U-B) magnitudes could be obtained for only 73 of these.

## Acknowledgements

This research was partially supported by the Indo-French Centre for the Promotion of Advanced Research / Centre Franco-Indien pour la Promotion de la Recherche Avancée, New-Delhi (India).

## References

Chevalier, C., Ilovaisky, S.A. 1991, *Astr. Astrophys. Suppl.*, **90**, 225.
Landolt, A.U. 1983, *Astr. J.*, **88**, 439.
Mohan, V., Paliwal,D.C., Mahara, H.S. 1991, *Bull. astr. Soc. India.*, **19**, 235.
Notni, P., Börngen, F., Ziener, R. 1979, *Astron. Nachr.*, **300**, 287.
Stetson, P.B. 1987, *Publ. astr. Soc. Pacific.*, **99**, 191.

**Caption to Figure**

Figure 1. Identification charts of the observed fields.

**Table 1.** Log of observations of CCD fields

| Field | Date of Observations | $\alpha$(2000) | $\delta$(2000) | Filters | Exposure VBU (in min) |
|---|---|---|---|---|---|
| *UPSO observations* | | | | | |
| Field-1 | 22-01-1993 | $9^h\ 21^m\ 30^s$ | $+50°\ 23'.5$ | UBV | 24,45,60 |
| Field-2 | 17-01-1993 | 9 21 39 | +47 51.5 | BV | 30,40 |
| Field-3 | 26-01-1993 | 9 22 00 | +48 45.5 | UBV | 30,45,90 |
| Field-4 | 28-12-1992 | 9 28 28 | +48 48.0 | UBV | 9,22,75 |
| Field-5 | 27-01-1993 | 9 28 32 | +47 20.0 | UBV | 10,20,45 |
| Field-6 | 24-01-1993 | 9 29 22 | +50 12.0 | UBV | 20,55,90 |
| Field-7 | 18-01-1993 | 9 32 24 | +48 15.5 | UBV | 18,40,90 |
| Field-8 | 20-01-1993 | 9 35 12 | +47 21.5 | UBV | 35,60,90 |
| Field-9 | 13-01-1991 | 9 37 46 | +48 43.5 | UBV | 10,20,30 |
| Field-10 | 02-01-1992 | 9 37 54 | +50 17.5 | UBV | 15,30,60 |
| Field-11 | 14-01-1991 | 9 40 15 | +48 38.5 | UBV | 15,20,30 |
| Field-12 | 19-04-1991 | 9 44 12 | +49 37.0 | UBV | 0.5,1,5 |
| | 23-12-1992 | | | UBV | 1,3,10 |
| Field-13 | 19-01-1993 | 9 48 02 | +47 26.0 | UBV | 30,45,90 |
| *Sub-portion of OHP fields Observed at UPSO* | | | | | |
| Field-14 | 24-12-1992 | 9 39 47 | +49 38.0 | UBV | 8,22,80 |
| Field-15 | 26-12-1992 | 9 44 56 | +49 42.0 | UBV | 10,30,90 |
| Field-16 | 21-12-1992 | 9 44 44 | +49 34.5 | UBV | 6,20,25 |
| Field-17 | 22-12-1992 | 9 47 27 | +50 04.0 | UBV | 10,25,45 |
| Field-18 | 23-12-1992 | 9 48 25 | +49 16.5 | UBV | 15,40,90 |
| *OHP observations* | | | | | |
| Field-14 | 4,5,6, 8-12-1991 | 9 39 56 | +49 36.0 | BV | 45,90 |
| Field-15 | 1,3,4-12-1991 | 9 44 54 | +49 42.0 | BV | 15,45 |
| Field-16 | 1,2,3, 4-12-1991 | 9 44 41 | +49 35.0 | BV | 35,60 |
| Field-17 | 30-11-1991 1,3,5-12-1991 | 9 47 36 | +50 05.0 | BV | 40,75 |
| Field-18 | 4,5-12-1991 | 9 48 32 | +49 13.5 | BV | 26,80 |

Table 2. Magnitudes and colours of observed stars

| Star no. | $\alpha(2000)$ | | | $\delta(2000)$ | | | V | B-V | U-B |
|---|---|---|---|---|---|---|---|---|---|
| | h | m | s | o | ′ | ″ | | | |
| *Field-1* | | | | | | | | | |
| 1 | 9 | 21 | 25.5 | +50 | 22 | 7 | 16.77 | 0.45 | -0.12 |
| 2 | 9 | 21 | 29.2 | +50 | 22 | 5 | 14.22 | 0.31 | 0.02 |
| 3 | 9 | 21 | 34.2 | +50 | 23 | 21 | 17.04 | 0.67 | -0.04 |
| 4 | 9 | 21 | 35.9 | +50 | 24 | 29 | 16.92 | 1.47 | * |
| | | | | | | | | | |
| *Field-2* | | | | | | | | | |
| 5 | 9 | 21 | 34.0 | +47 | 52 | 58 | 16.48 | 0.51 | * |
| 6 | 9 | 21 | 36.4 | +47 | 51 | 39 | 18.85 | 1.38 | * |
| 7 | 9 | 21 | 38.4 | +47 | 50 | 33 | 19.35 | 1.08 | * |
| 8 | 9 | 21 | 39.6 | +47 | 53 | 21 | 17.95 | 1.18 | * |
| 9 | 9 | 21 | 40.3 | +47 | 53 | 14 | 17.15 | 0.96 | * |
| 10 | 9 | 21 | 40.7 | +47 | 51 | 2 | 18.27 | 0.70 | * |
| 11 | 9 | 21 | 42.6 | +47 | 52 | 54 | 20.24 | 0.70 | * |
| 12 | 9 | 21 | 43.5 | +47 | 53 | 23 | 17.28 | 1.07 | * |
| | | | | | | | | | |
| *Field-3* | | | | | | | | | |
| 13 | 9 | 21 | 54.8 | +48 | 44 | 24 | 16.85 | 0.93 | 0.87 |
| 14 | 9 | 21 | 54.9 | +48 | 46 | 38 | 17.47 | 0.65 | -0.18 |
| 15 | 9 | 21 | 56.7 | +48 | 46 | 33 | 16.94 | 1.43 | * |
| 16 | 9 | 22 | 1.1 | +48 | 46 | 41 | 14.92 | 0.71 | 0.24 |
| 17 | 9 | 22 | 2.4 | +48 | 44 | 15 | 15.12 | 1.17 | 1.09 |
| 18 | 9 | 22 | 3.9 | +48 | 44 | 56 | 14.34 | 0.65 | 0.15 |
| | | | | | | | | | |
| *Field-4* | | | | | | | | | |
| 19 | 9 | 28 | 20.7 | +48 | 48 | 2 | 14.73 | 1.04 | 0.78 |
| 20 | 9 | 28 | 22.8 | +48 | 47 | 4 | 11.05 | 0.72 | 0.27 |
| 21 | 9 | 28 | 25.9 | +48 | 47 | 53 | 15.12 | 1.56 | * |
| 22 | 9 | 28 | 32.1 | +48 | 48 | 25 | 14.39 | 0.76 | 0.45 |
| 23 | 9 | 28 | 36.0 | +48 | 48 | 52 | 11.57 | 0.59 | 0.04 |
| | | | | | | | | | |
| *Field-5* | | | | | | | | | |
| 24 | 9 | 28 | 30.5 | +47 | 19 | 9 | 13.78 | 0.71 | 0.24 |
| 25 | 9 | 28 | 33.8 | +47 | 19 | 21 | 15.70 | 0.73 | 0.27 |
| 26 | 9 | 28 | 34.6 | +47 | 21 | 5 | 12.11 | 0.92 | 0.39 |
| 27 | 9 | 28 | 35.1 | +47 | 18 | 33 | 15.85 | 0.66 | 0.08 |
| 28 | 9 | 28 | 35.7 | +47 | 21 | 34 | 17.26 | 1.09 | * |

Table 2..... contd.

| Star no. | α(2000) | | | δ(2000) | | | V | B-V | U-B |
|---|---|---|---|---|---|---|---|---|---|
| | h | m | s | o | ' | " | | | |
| *Field-6* | | | | | | | | | |
| 29 | 9 | 29 | 15.5 | +50 | 13 | 35 | 16.76 | 0.31 | -0.64 |
| 30 | 9 | 29 | 16.4 | +50 | 10 | 45 | 16.60 | 0.55 | 0.23 |
| 31 | 9 | 29 | 23.5 | +50 | 11 | 27 | 14.37 | 1.19 | 1.23 |
| 32 | 9 | 29 | 25.6 | +50 | 12 | 14 | 16.98 | 0.48 | -0.03 |
| 33 | 9 | 29 | 25.5 | +50 | 12 | 58 | 14.49 | 0.79 | 0.58 |
| *Field-7* | | | | | | | | | |
| 34 | 9 | 32 | 18.5 | +48 | 17 | 14 | 17.08 | 1.33 | * |
| 35 | 9 | 32 | 19.2 | +48 | 16 | 32 | 16.84 | 0.54 | 0.04 |
| 36 | 9 | 32 | 19.8 | +48 | 17 | 33 | 19.27 | 1.38 | * |
| 37 | 9 | 32 | 22.9 | +48 | 15 | 55 | 17.50 | 0.40 | -0.13 |
| 38 | 9 | 32 | 23.1 | +48 | 14 | 30 | 15.24 | 0.60 | 0.42 |
| 39 | 9 | 32 | 23.8 | +48 | 15 | 50 | 19.51 | 0.71 | * |
| 40 | 9 | 32 | 28.2 | +48 | 14 | 25 | 13.66 | 0.55 | 0.34 |
| 41 | 9 | 32 | 28.8 | +48 | 14 | 19 | 18.92 | 0.37 | * |
| 42 | 9 | 32 | 30.2 | +48 | 15 | 58 | 15.53 | 1.18 | 1.39 |
| *Field-8* | | | | | | | | | |
| 43 | 9 | 35 | 7.1 | +47 | 23 | 14 | 13.75 | 1.10 | 1.12 |
| 44 | 9 | 35 | 7.5 | +47 | 21 | 18 | 16.35 | 0.60 | 0.08 |
| 45 | 9 | 35 | 10.3 | +47 | 21 | 41 | 15.83 | 0.85 | 0.40 |
| 46 | 9 | 35 | 13.0 | +47 | 20 | 6 | 15.22 | 0.99 | 0.79 |
| 47 | 9 | 35 | 13.6 | +47 | 20 | 56 | 19.45 | 0.41 | * |
| 48 | 9 | 35 | 15.6 | +47 | 22 | 22 | 18.98 | 0.82 | * |
| 49 | 9 | 35 | 16.1 | +47 | 21 | 39 | 15.37 | 0.53 | -0.03 |
| *Field-9* | | | | | | | | | |
| 50 | 9 | 37 | 43.6 | +48 | 44 | 31 | 16.06 | 0.67 | 0.19 |
| 51 | 9 | 37 | 47.3 | +48 | 42 | 36 | 15.65 | 0.54 | 0.00 |
| 52 | 9 | 37 | 48.9 | +48 | 43 | 32 | 13.92 | 0.71 | 0.41 |
| *Field-10* | | | | | | | | | |
| 53 | 9 | 37 | 48.2 | +50 | 18 | 27 | 14.55 | 0.67 | 0.58 |
| 54 | 9 | 37 | 48.3 | +50 | 17 | 27 | 15.45 | 0.68 | 0.31 |
| 55 | 9 | 37 | 49.7 | +50 | 18 | 42 | 18.68 | 0.55 | * |
| 56 | 9 | 37 | 57.4 | +50 | 17 | 46 | 15.69 | 0.81 | 0.30 |
| 57 | 9 | 37 | 59.7 | +50 | 16 | 53 | 15.57 | 0.85 | 0.48 |
| 58 | 9 | 38 | 1.7 | +50 | 17 | 26 | 16.01 | 1.00 | 0.72 |

**Table 2..... contd.**

| Star no. | α(2000) | | | δ(2000) | | | V | B-V | U-B |
|---|---|---|---|---|---|---|---|---|---|
| | h | m | s | o | ′ | ″ | | | |
| *Field-11* | | | | | | | | | |
| 59 | 9 | 40 | 12.1 | +48 | 37 | 25 | 14.04 | 0.57 | 0.00 |
| 60 | 9 | 40 | 14.7 | +48 | 38 | 54 | 12.60 | 0.48 | -0.07 |
| 61 | 9 | 40 | 17.1 | +48 | 39 | 31 | 12.08 | 0.65 | 0.15 |
| 62 | 9 | 40 | 18.1 | +48 | 37 | 54 | 15.82 | 0.77 | * |
| *Field-12* | | | | | | | | | |
| 63 | 9 | 44 | 5.0 | +49 | 36 | 19 | 14.06 | 0.48 | -0.05 |
| 64 | 9 | 44 | 11.3 | +49 | 38 | 9 | 12.14 | 0.59 | 0.06 |
| 65 | 9 | 44 | 13.7 | +49 | 35 | 31 | 14.70 | 0.52 | -0.20 |
| 66 | 9 | 44 | 15.0 | +49 | 37 | 7 | 11.15 | 0.94 | 0.66 |
| *Field-13* | | | | | | | | | |
| 67 | 9 | 47 | 56.4 | +47 | 25 | 4 | 15.56 | 0.78 | 0.12 |
| 68 | 9 | 47 | 57.7 | +47 | 24 | 33 | 17.90 | 0.41 | -0.44 |
| 69 | 9 | 47 | 59.3 | +47 | 26 | 0 | 15.78 | 1.164 | 0.78 |
| 70 | 9 | 48 | 5.3 | +47 | 27 | 25 | 15.11 | 0.85 | 0.13 |
| 71 | 9 | 48 | 6.3 | +47 | 24 | 43 | 17.98 | 1.63 | * |
| 72 | 9 | 48 | 8.2 | +47 | 26 | 54 | 16.43 | 1.13 | 0.73 |
| *Field-14* | | | | | | | | | |
| 73 | 9 | 39 | 44.2 | +49 | 34 | 15 | 17.48 | 0.51 | * |
| 74 | 9 | 39 | 45.2 | +49 | 36 | 42 | 12.46 | 0.44 | -0.22 |
| 75 | 9 | 39 | 45.8 | +49 | 39 | 14 | 19.61 | 0.46 | * |
| 76 | 9 | 39 | 47.4 | +49 | 36 | 6 | 11.18 | 0.18 | * |
| 77 | 9 | 39 | 49.4 | +49 | 38 | 59 | 14.32 | 0.95 | 0.63 |
| 78 | 9 | 39 | 50.1 | +49 | 38 | 5 | 14.53 | 0.66 | 0.01 |
| 79 | 9 | 39 | 50.4 | +49 | 34 | 30 | 15.97 | 0.74 | * |
| 80 | 9 | 39 | 50.4 | +49 | 36 | 22 | 17.99 | 1.33 | * |
| 81 | 9 | 39 | 51.0 | +49 | 41 | 4 | 18.30 | 1.86 | * |
| 82 | 9 | 39 | 52.2 | +49 | 37 | 1 | 17.43 | 1.32 | * |
| 83 | 9 | 39 | 53.4 | +49 | 39 | 24 | 19.82 | 0.88 | * |
| 84 | 9 | 39 | 53.3 | +49 | 35 | 18 | 19.16 | 1.63 | * |
| 85 | 9 | 39 | 53.5 | +49 | 36 | 24 | 20.39 | 0.64 | * |
| 86 | 9 | 39 | 55.0 | +49 | 34 | 54 | 19.24 | 0.80 | * |
| 87 | 9 | 39 | 56.4 | +49 | 36 | 40 | 18.68 | 0.57 | * |
| 88 | 9 | 39 | 59.2 | +49 | 34 | 50 | 19.24 | 1.44 | * |
| 89 | 9 | 40 | 0.0 | +49 | 34 | 8 | 19.34 | 1.40 | * |
| 90 | 9 | 39 | 59.8 | +49 | 38 | 7 | 18.50 | 0.61 | * |

Table 2..... contd.

| Star no. | α(2000) | | | δ(2000) | | | V | B-V | U-B |
|---|---|---|---|---|---|---|---|---|---|
| | h | m | s | o | ′ | ″ | | | |
| 91 | 9 | 40 | 0.7 | +49 | 34 | 52 | 19.53 | 0.52 | * |
| 92 | 9 | 40 | 2.5 | +49 | 37 | 42 | 14.45 | 0.68 | * |
| 93 | 9 | 40 | 4.6 | +49 | 37 | 16 | 15.28 | 0.72 | * |
| 94 | 9 | 40 | 7.7 | +49 | 35 | 34 | 18.59 | 1.75 | * |
| 95 | 9 | 40 | 9.0 | +49 | 37 | 1 | 17.07 | 0.66 | * |
| 96 | 9 | 40 | 9.1 | +49 | 39 | 57 | 20.61 | 0.52 | * |
| *Field-15* | | | | | | | | | |
| 97 | 9 | 44 | 42.0 | +49 | 39 | 42 | 19.43 | 0.63 | * |
| 98 | 9 | 44 | 42.2 | +49 | 44 | 56 | 20.15 | 1.33 | * |
| 99 | 9 | 44 | 43.2 | +49 | 42 | 44 | 20.09 | 1.23 | * |
| 100 | 9 | 44 | 45.3 | +49 | 40 | 8 | 14.44 | 0.47 | * |
| 101 | 9 | 44 | 45.5 | +49 | 42 | 29 | 18.44 | 1.57 | * |
| 102 | 9 | 44 | 46.3 | +49 | 42 | 52 | 19.50 | 1.58 | * |
| 103 | 9 | 44 | 49.0 | +49 | 41 | 50 | 20.77 | 0.34 | * |
| 104 | 9 | 44 | 50.8 | +49 | 40 | 48 | 16.19 | 0.45 | -0.07 |
| 105 | 9 | 44 | 51.4 | +49 | 42 | 42 | 19.80 | 1.40 | * |
| 106 | 9 | 44 | 52.3 | +49 | 45 | 32 | 15.27 | 0.72 | * |
| 107 | 9 | 44 | 56.4 | +49 | 40 | 41 | 15.58 | 0.63 | 0.29 |
| 108 | 9 | 44 | 57.0 | +49 | 44 | 31 | 20.01 | 1.71 | * |
| 109 | 9 | 44 | 58.6 | +49 | 41 | 18 | 13.69 | 0.71 | 0.31 |
| 110 | 9 | 45 | 0.1 | +49 | 42 | 52 | 12.98 | 0.88 | 0.75 |
| 111 | 9 | 45 | 1.9 | +49 | 42 | 15 | 20.43 | 0.93 | * |
| 112 | 9 | 45 | 1.9 | +49 | 39 | 32 | 18.32 | 0.95 | * |
| 113 | 9 | 45 | 2.4 | +49 | 39 | 19 | 19.67 | 1.62 | * |
| 114 | 9 | 45 | 2.6 | +49 | 39 | 54 | 18.85 | 0.91 | * |
| 115 | 9 | 45 | 5.0 | +49 | 42 | 8 | 19.32 | 1.33 | * |
| 116 | 9 | 45 | 7.0 | +49 | 42 | 27 | 20.39 | 0.31 | * |
| *Field-16* | | | | | | | | | |
| 117 | 9 | 44 | 29.6 | +49 | 35 | 29 | 20.44 | 0.55 | * |
| 118 | 9 | 44 | 31.1 | +49 | 35 | 46 | 20.68 | 1.13 | * |
| 119 | 9 | 44 | 31.6 | +49 | 34 | 43 | 20.28 | 1.43 | * |
| 120 | 9 | 44 | 31.4 | +49 | 33 | 13 | 20.47 | 0.97 | * |
| 121 | 9 | 44 | 32.1 | +49 | 35 | 29 | 16.69 | 0.95 | * |
| 122 | 9 | 44 | 32.6 | +49 | 32 | 36 | 18.43 | 0.53 | * |
| 123 | 9 | 44 | 33.5 | +49 | 37 | 22 | 17.97 | 0.63 | * |
| 124 | 9 | 44 | 34.8 | +49 | 35 | 44 | 18.24 | 1.41 | * |
| 125 | 9 | 44 | 35.9 | +49 | 35 | 17 | 18.85 | 1.21 | * |

Table 2..... contd.

| Star no. | α(2000) | | | δ(2000) | | | V | B-V | U-B |
|---|---|---|---|---|---|---|---|---|---|
| | h | m | s | o | ′ | ″ | | | |
| 126 | 9 | 44 | 37.8 | +49 | 38 | 15 | 20.23 | 0.20 | * |
| 127 | 9 | 44 | 37.7 | +49 | 32 | 19 | 19.66 | 1.26 | * |
| 128 | 9 | 44 | 37.9 | +49 | 33 | 16 | 19.30 | 0.90 | * |
| 129 | 9 | 44 | 38.5 | +49 | 33 | 34 | 13.83 | 0.60 | 0.11 |
| 130 | 9 | 44 | 39.9 | +49 | 31 | 59 | 19.10 | 1.51 | * |
| 131 | 9 | 44 | 40.6 | +49 | 36 | 3 | 14.51 | 0.65 | -0.03 |
| 132 | 9 | 44 | 40.8 | +49 | 32 | 0 | 18.35 | 1.43 | * |
| 133 | 9 | 44 | 41.1 | +49 | 33 | 18 | 17.21 | 0.82 | 0.36 |
| 134 | 9 | 44 | 42.7 | +49 | 32 | 43 | 20.59 | 0.32 | * |
| 135 | 9 | 44 | 43.7 | +49 | 36 | 56 | 18.08 | 1.27 | * |
| 136 | 9 | 44 | 44.8 | +49 | 35 | 47 | 16.13 | 0.75 | 0.18 |
| 137 | 9 | 44 | 45.0 | +49 | 32 | 13 | 18.80 | 0.73 | * |
| 138 | 9 | 44 | 48.8 | +49 | 35 | 37 | 19.46 | 0.69 | * |
| 139 | 9 | 44 | 49.5 | +49 | 33 | 53 | 17.95 | 1.64 | * |
| 140 | 9 | 44 | 49.2 | +49 | 37 | 10 | 19.25 | 1.55 | * |
| 141 | 9 | 44 | 50.2 | +49 | 35 | 47 | 12.21 | 0.46 | 0.00 |
| 142 | 9 | 44 | 50.8 | +49 | 32 | 2 | 18.25 | 1.30 | * |
| 143 | 9 | 44 | 51.7 | +49 | 35 | 57 | 18.16 | 1.27 | * |
| *Field-17* | | | | | | | | | |
| 144 | 9 | 47 | 21.1 | +50 | 4 | 23 | 19.46 | 1.30 | * |
| 145 | 9 | 47 | 21.5 | +50 | 5 | 6 | 15.36 | 0.55 | 0.23 |
| 146 | 9 | 47 | 22.4 | +50 | 4 | 23 | 17.83 | 0.69 | * |
| 147 | 9 | 47 | 23.3 | +50 | 5 | 28 | 19.64 | 1.23 | * |
| 148 | 9 | 47 | 26.0 | +50 | 4 | 0 | 19.26 | 1.51 | * |
| 149 | 9 | 47 | 26.5 | +50 | 7 | 51 | 20.28 | 0.83 | * |
| 150 | 9 | 47 | 26.4 | +50 | 7 | 26 | 18.09 | 1.25 | * |
| 151 | 9 | 47 | 28.7 | +50 | 5 | 0 | 15.88 | 0.91 | 0.62 |
| 152 | 9 | 47 | 29.2 | +50 | 5 | 33 | 14.33 | 0.64 | 0.12 |
| 153 | 9 | 47 | 29.7 | +50 | 4 | 52 | 14.21 | 0.79 | 0.39 |
| 154 | 9 | 47 | 30.2 | +50 | 5 | 3 | 20.79 | 0.93 | * |
| 155 | 9 | 47 | 31.1 | +50 | 8 | 48 | 18.68 | 1.07 | * |
| 156 | 9 | 47 | 31.6 | +50 | 5 | 17 | 19.35 | 0.99 | * |
| 157 | 9 | 47 | 31.8 | +50 | 5 | 38 | 15.05 | 0.83 | 0.51 |
| 158 | 9 | 47 | 32.8 | +50 | 3 | 3 | 14.01 | 0.63 | 0.23 |
| 159 | 9 | 47 | 33.1 | +50 | 2 | 41 | 13.19 | 0.56 | -0.04 |
| 160 | 9 | 47 | 33.3 | +50 | 4 | 3 | 18.73 | 1.42 | * |

Table 2..... contd.

| Star no. | α(2000) | | | δ(2000) | | | V | B-V | U-B |
|---|---|---|---|---|---|---|---|---|---|
| | h | m | s | ° | ′ | ″ | | | |
| 161 | 9 | 47 | 33.6 | +50 | 7 | 37 | 16.97 | 0.75 | * |
| 162 | 9 | 47 | 34.3 | +50 | 7 | 52 | 17.98 | 0.75 | * |
| 163 | 9 | 47 | 34.2 | +50 | 3 | 38 | 18.45 | 1.31 | * |
| 164 | 9 | 47 | 35.2 | +50 | 3 | 42 | 19.20 | 1.45 | * |
| 165 | 9 | 47 | 35.9 | +50 | 3 | 14 | 19.89 | 1.41 | * |
| 166 | 9 | 47 | 35.9 | +50 | 4 | 31 | 19.44 | 1.30 | * |
| 167 | 9 | 47 | 36.2 | +50 | 3 | 24 | 18.96 | 1.44 | * |
| 168 | 9 | 47 | 37.6 | +50 | 6 | 21 | 16.77 | 0.51 | * |
| 169 | 9 | 47 | 38.2 | +50 | 3 | 18 | 17.65 | 0.95 | * |
| 170 | 9 | 47 | 38.4 | +50 | 5 | 5 | 19.70 | 0.19 | * |
| 171 | 9 | 47 | 38.5 | +50 | 2 | 46 | 20.72 | 0.88 | * |
| 172 | 9 | 47 | 39.0 | +50 | 7 | 46 | 18.03 | 0.83 | * |
| 173 | 9 | 47 | 39.3 | +50 | 2 | 49 | 19.34 | 0.99 | * |
| 174 | 9 | 47 | 39.3 | +50 | 4 | 42 | 20.14 | 1.29 | * |
| 175 | 9 | 47 | 41.5 | +50 | 3 | 14 | 19.31 | 1.70 | * |
| 176 | 9 | 47 | 41.3 | +50 | 6 | 55 | 18.11 | 1.08 | * |
| 177 | 9 | 47 | 42.4 | +50 | 8 | 16 | 20.28 | 0.32 | * |
| 178 | 9 | 47 | 42.7 | +50 | 3 | 53 | 15.28 | 0.95 | * |
| 179 | 9 | 47 | 45.6 | +50 | 7 | 44 | 18.73 | 1.54 | * |
| 180 | 9 | 47 | 46.4 | +50 | 4 | 18 | 20.15 | 1.59 | * |
| 181 | 9 | 47 | 46.2 | +50 | 7 | 27 | 20.35 | 0.85 | * |
| 182 | 9 | 47 | 46.4 | +50 | 5 | 58 | 20.59 | 0.44 | * |
| 183 | 9 | 47 | 47.0 | +50 | 5 | 46 | 19.95 | 0.85 | * |
| 184 | 9 | 47 | 50.5 | +50 | 7 | 45 | 19.91 | 0.78 | * |
| *Field-18* | | | | | | | | | |
| 186 | 9 | 48 | 19.1 | +49 | 15 | 19 | 16.74 | 0.78 | * |
| 187 | 9 | 48 | 19.4 | +49 | 15 | 51 | 17.90 | 0.97 | * |
| 188 | 9 | 48 | 20.3 | +49 | 13 | 28 | 19.21 | 0.29 | * |
| 189 | 9 | 48 | 21.3 | +49 | 16 | 42 | 19.88 | 0.68 | * |
| 190 | 9 | 48 | 21.6 | +49 | 15 | 58 | 17.89 | 0.70 | * |
| 191 | 9 | 48 | 21.8 | +49 | 14 | 45 | 20.18 | 1.00 | * |
| 192 | 9 | 48 | 22.7 | +49 | 12 | 58 | 16.73 | 0.98 | * |
| 193 | 9 | 48 | 22.8 | +49 | 13 | 55 | 18.03 | 1.10 | * |
| 194 | 9 | 48 | 23.7 | +49 | 15 | 11 | 16.15 | 0.64 | 0.18 |
| 195 | 9 | 48 | 24.6 | +49 | 16 | 59 | 14.82 | 0.64 | 0.16 |
| 196 | 9 | 48 | 24.5 | +49 | 13 | 29 | 18.56 | 1.35 | * |
| 197 | 9 | 48 | 24.4 | +49 | 12 | 58 | 18.90 | 1.20 | * |
| 198 | 9 | 48 | 24.9 | +49 | 15 | 51 | 20.35 | 0.77 | * |

Table 2..... contd.

| Star no. | α(2000) | | | δ(2000) | | | V | B-V | U-B |
|---|---|---|---|---|---|---|---|---|---|
| | h | m | s | o | ' | " | | | |
| 199 | 9 | 48 | 25.8 | +49 | 10 | 38 | 12.62 | 0.60 | * |
| 200 | 9 | 48 | 27.7 | +49 | 15 | 13 | 15.75 | 1.03 | 1.00 |
| 201 | 9 | 48 | 28.3 | +49 | 13 | 27 | 20.48 | 1.17 | * |
| 202 | 9 | 48 | 29.5 | +49 | 12 | 56 | 17.47 | 0.96 | * |
| 203 | 9 | 48 | 29.9 | +49 | 11 | 39 | 15.36 | 0.55 | * |
| 204 | 9 | 48 | 31.7 | +49 | 11 | 40 | 19.67 | 1.18 | * |
| 205 | 9 | 48 | 33.6 | +49 | 14 | 19 | 13.40 | 0.56 | * |
| 206 | 9 | 48 | 33.8 | +49 | 11 | 46 | 19.74 | 1.09 | * |
| 207 | 9 | 48 | 34.3 | +49 | 11 | 41 | 19.65 | 0.65 | * |
| 208 | 9 | 48 | 35.8 | +49 | 12 | 14 | 19.64 | 0.69 | * |
| 209 | 9 | 48 | 38.4 | +49 | 17 | 18 | 19.19 | 0.56 | * |
| 210 | 9 | 48 | 39.1 | +49 | 13 | 32 | 16.90 | 0.53 | * |
| 211 | 9 | 48 | 42.7 | +49 | 10 | 34 | 15.79 | 0.65 | * |
| 212 | 9 | 48 | 43.5 | +49 | 17 | 6 | 13.26 | 0.58 | * |
| 213 | 9 | 48 | 44.3 | +49 | 11 | 36 | 16.99 | 0.90 | * |
| 214 | 9 | 48 | 44.7 | +49 | 15 | 15 | 18.12 | 0.56 | * |